\documentstyle[12pt,aasms4]{article}

\begin{document}

\def\mh{[M/H]}
\def\mk{M_K}
\def\mv{M_V}
\def\mi{M_I}
\def\MH{{\rm [M/H]}}
\def\mvol{M_\odot pc^{-3}}
\def\msol{M_\odot}
\def\zsol{Z_\odot}
\def\te{T_{eff}}
\def\simgr{\,\hbox{\hbox{$ > $}\kern -0.8em \lower 1.0ex\hbox{$\sim$}}\,}
\def\simle{\,\hbox{\hbox{$ < $}\kern -0.8em \lower 1.0ex\hbox{$\sim$}}\,}
\def\wig#1{\mathrel{\hbox{\hbox to 0pt{%
          \lower.5ex\hbox{$\sim$}\hss}\raise.4ex\hbox{$#1$}}}}
\newcommand\etal{{\it et al.}}

\title{\Large \bf The Galactic disk mass-budget : I. stellar mass-function and density.}

\author{Gilles Chabrier}
\affil{
Ecole Normale Superieure de Lyon,\\
C.R.A.L. (UMR 5574 CNRS),
69364 Lyon Cedex 07, France\\
and\\
Visiting Miller Professor, Dpt. of Astronomy, University of Berkeley,\\
Berkeley, CA 94720}

\authoremail{chabrier@ens-lyon.fr}

\begin{center}

\bigskip
\end{center}
\bigskip
\bigskip

\begin{abstract}

In this paper, we use the general theory worked out within the past few years for the structure and the evolution of low-mass stars to derive the stellar mass-function in the Galactic disk down to
the vicinity of the hydrogen-burning limit, from the observed nearby luminosity functions. The accuracy of the mass-magnitude relationships derived from the afore-mentioned theory is examined by comparison with recent, accurate observational relationships in the M-dwarf domain.
The mass function is shown to flatten out below $\sim 1\,\msol$ but to keep rising down to the bottom of the main sequence. Combining the present determination below 1 $\msol$ and Scalo's (1986) mass function for larger masses,
we show that the mass function is well described over the entire stellar mass range, from $\sim 100\,\msol$ to $\sim 0.1\,\msol$, by three functional forms, namely a two-segment power-law, a log-normal form or an exponential form, all normalized to the Hipparcos sample at 0.8 $\msol$.

Integration of this mass function yields a reasonably accurate census of the entire stellar population in the Galactic disk, and its volume and surface mass-density.
\end{abstract}


\section{Introduction}

A correct determination of the Galactic mass function (MF) over the entire stellar domain is essential for several reasons. First it yields the complete census of the stellar population in the Galaxy and thus its contribution to the Galactic baryonic content and mass-to-light ratio. Second, the normalization and the slope of the stellar MF near the hydrogen-burning limit provides the boundary condition to infer the brown dwarf MF, assuming the MF is continuous at the H-burning limit. Third, the determination of the global
stellar plus substellar MF provides an essential diagnostic to understand the formation of star-like objects.

Star counts, which a few years ago still relied essentially on giants and sun-like stars, i.e. objects
with mass $m\simgr 1\,\msol$, now include information on the M-dwarf population in the Galaxy. As discussed below, modern observations now probe the M-dwarf stellar distribution
down to the bottom of the main sequence (MS), close to the hydrogen-burning 
limit. Moreover, we know that this distribution extends into the substellar regime since over a hundred field brown dwarfs have now been discovered in the solar neighborhood.

In the present paper, we define the initial mass function (IMF) $\xi$ as the number of stars $N$, or the stellar number-density per cubic parsec $n$, formed initially per mass interval $[m,m+dm]$:

\begin{eqnarray}
\xi(m)={dN\over dm}
\end{eqnarray}

The MF can also be described as the number or the density of stars per interval of $\log m$, as defined originally by Salpeter (1955),
with the straightforward relation:

\begin{eqnarray}
\xi\log (m)={dN\over d\log m}=({\rm Ln} \,10)\,m\,\xi(m)
\end{eqnarray}

The quest for the determination of the stellar IMF takes root in the seminal paper by Salpeter (1955) and in the following major contributions by Miller \& Scalo (1979), Scalo (1986), and Kroupa, Tout \& Gilmore (1991, 1993). All these papers agree for an IMF reasonably well described by a power-law form $\xi(m) \propto m^{-\alpha}$ with an exponent $\alpha \simeq 2.35$ for stars with $m\simgr 1 \,\msol$, the so-called Salpeter IMF, although the more detailed analysis of Scalo (1986) suggests a steeper slope, with $\alpha \simeq 2.7$. A recent analysis by Kroupa (2001) suggests $\alpha \sim 2.3$ but, as noted by this author, the previous Scalo value is recovered if one assumes a standard fraction of observationally unresolved binaries for these massive stars. On the other hand, the IMF seems to flatten significantly at the low-mass end, as suggested initially by
Miller \& Scalo (1979). Indeed, the detailed analysis of Kroupa et al. (1993) suggests a significant flattening of the IMF below
$\sim 0.5\,\msol$, with $\alpha \simeq 1.3$,
although the question of whether the IMF keeps rising down to the H-burning limit or decreases below a certain mass around $\sim 0.2 \,\msol$ remains unclear. In fact the very functional form of the IMF in the M-dwarf range, power-law, log-normal or other form, is presently still undetermined.
It is important to note, in the present context, that the approach of Kroupa and collaborators, who
first addressed this problem in the M-dwarf range, is essentially empirical, in the sense that these authors used
an empirical mass-luminosity relationship derived from observations
of nearby binaries (Popper, 1980).

The aim of the present paper is to {\it demonstrate} those results, in deriving a stellar MF from a
consistent stellar evolution {\it theory}, and to explore the possibility of various functional forms for the IMF in the low-mass range. As mentioned below, this theory describes accurately the very mechanical and thermal properties of M-dwarf stars, and yields reliable mass-magnitude relationships, allowing age and metallicity determinations.
The present paper presents the derivation of the stellar IMF for the Galactic disk
over the entire stellar mass range, with particular emphasis on the low-mass part of the distribution ($m< 1 \,\msol$). The accuracy of the stellar models, more particularly of the mass-magnitude relationships, is examined in \S2 by comparison with the most recent
observationally determined relationships in different optical and infrared passbands. In \S3, we examine the different M-dwarf luminosity functions presently available for probing
the disc low-mass star distribution. The mass functions are derived in \S4. In this section, we carefully examine the uncertainties due to the mass-magnitude relationship and we derive analytical parametrizations of the MF. The inferred stellar mass budget is determined in \S5 while section 6 is devoted to the conclusion. The extension of these calculations into the brown dwarf regime, where age becomes an extra degree of freedom besides mass, will be presented in a subsequent paper.

\section{The mass-magnitude relationships}

The key ingredient in the determination of the MF is the mass-magnitude relationship (MMR). As too rarely stressed, the MMR is the cornerstone to transform the observable quantity, the luminosity function (LF) $\phi=dN/dM$, i.e. the number of stars $N$ per absolute magnitude interval $[M,M+dM]$, into a MF.
The derivation of observable MMRs for M-dwarfs is a formidable task for several reasons. First of all, we
need to determine the absolute magnitudes of the objects, which implies accurate parallaxes,
and thus nearby distances.
Second, M-dwarfs are by definition intrinsically faint, in particular near the bottom of the MS, where the
determination of the MMR is most crucial since the luminosity drops by orders of magnitude
with the mass when approaching the hydrogen-burning limit.
At last, the determination of the mass requires binary systems, which reduces the already weak statistics
by about a factor 2, and dynamical information, i.e. a long enough time basis. All these difficulties render
the determination of an observed M-dwarf MMR a formidable challenge for astronomers.
A first compilation of mass-luminosity data in the M-dwarf domain was published by Popper (1980), and was subsequently
extended by Henry \& McCarthy (1993). These authors
used speckle interferometry to
obtain MMRs in the V-, J-, H- and K-bands. The determination of the V-magnitude of these objects has been improved recently
by using the HST (Henry et al., 1999), reducing appreciably the uncertainty in the $m$-$\mv$ relation.
The observationally-determined MMRs were fitted by simple polynomial
expressions. As mentioned by the authors themselves, these fits are to be taken with great caution and must be used as guidelines only,
for they do not take into account age and metallicity-dispersion, characteristic of the young disk/old disk population.
The Henry \& McCarthy (1993) sample has been improved significantly recently by Delfosse et al. (2000). These authors
combined adaptative optics images and accurate radial velocities to determine the mass of about 20 objects between
$\sim 0.6$ and $\sim 0.09$ $\msol$ in the $V$-, $J$-, $H$- and $K$-bands. The masses of the Delfosse et al. sample are obtained with accuracies of 0.2 to 5\% and the distance is determined also with high precision.

Although such data provides extremely useful empirical MMRs, the MMR is generally derived from theoretical models aimed at describing as accurately as possible the structure and the evolution of the objects under study. Indeed the shape of the MMR reflects the very mechanical and thermal properties of these objects (see Chabrier \& Baraffe, 2000 for a recent review). The accuracy of the stellar models and of the derived theoretical MMRs is thus a crucial issue since the transformation of the LF into the MF involves the derivative of the MMR:

\begin{eqnarray}
 {dN\over dm}(m)=({dN\over dM_\lambda(m)})\times ({dm\over dM_\lambda(m)})^{-1}
\end{eqnarray}

\noindent where $N$ is the number of stars, $m$ the mass and $M_\lambda$ the absolute magnitude in a given passband.

For a claimed MF determination to retain any degree of reliability, it must rely on MMRs, and thus on stellar models, which have been
demonstrated to accurately reproduce observational data in all available observational diagrams, color-color, color-magnitude, mass-spectral type and most importantly mass-magnitude. This is mandatory for any MF determination. Only recently has a generation of low-mass stars (LMS) and brown dwarfs (BD) evolutionary models emerged which relies on consistent calculations of the internal and atmospheric properties of these objects, with no
adjustable parameter\footnote{Recall that below $\sim 0.7\,\msol$ the objects are dominantly or even entirely
convective so that variations of the mixing length parameter is inconsequential on the evolution (see Chabrier \& Baraffe, 1997 for a
detailed discussion)} and thus provide consistent mass-age-magnitude-color relationships. The physics of these models and extensive comparisons with observations in different observational diagrams have been discussed in various papers (Chabrier \& Baraffe 1997; Baraffe et al 1997, 1998; Chabrier et al., 2000; Chabrier \& Baraffe, 2000 and references therein).
Although improvement is still needed for
a complete agreement with all the observed magnitude-color diagrams (see Baraffe et al., 1998, hereafter BCAH98 and discussion in \S4.1), these
models improve significantly the situation, in particular in the infrared, and reach now quantitative agreement with the observations.
Figure 3 of Delfosse et al. (2000) shows a comparison of the MMRs issued from the afore-mentioned general theory with the Delfosse et al. (2000) recent data in the $V$, $J$, $H$ and $K$-bands. The agreement in the $J$, $H$ and $K$-bands is essentially perfect ($<$ 1$\sigma$).
The situation is not as satisfactory in the V-band, with a systematic
offset between theory and observation below $\sim 0.3\,\msol$, $\mv \simgr 12$.
The consequence of such an uncertainty on the derivation of the MF will be examined carefully in \S4.1. 

This general excellent agreement between theory and observation brings confidence in the description of the mechanical and thermal properties of low-mass stars down to the vicinity of the H-burning limit. It allows a more detailed study of the effects of age and metallicity on
the MMR.

\indent $\bullet$ age effect : objects with $m\simgr 0.1\,\msol$ reach the MS after $\sim 0.7$ Gyr (see e.g. Table 1 of Chabrier \& Baraffe, 2000) and,
for $m\simle 0.8\,\msol$, stay there for a Hubble time,
so that their position is fixed in the mass-luminosity diagram. Younger objects in this mass-range will still be on their pre-MS contraction
phase and will be more luminous than main sequence objects with the same mass (see Figure 3 of BCAH98).
Assuming a constant stellar formation rate and an age for the Galactic disk $t_D\approx 10$ Gyr, the fraction
of such young objects in the local population, however, should represent at most a few percents.

\indent $\bullet$ metallicity effect :
The Hipparcos color-magnitude diagram
indicates that $\sim 90\%$ of disk stars have abundances within $\pm 0.2$ dex of the solar value (Reid, 1999) so the spread of metallicity in the solar neighborhood should not affect
significantly the derivation of the MF through the MMR.
The consequence of variations of the metallicity over a range $\Delta\mh= \pm 0.5$ on the MF is examined in M\'era, Chabrier \& Baraffe (1996). A lower metallicity yields a steeper slope since the lower the metallicity the brighter the star for a given mass or, conversely, the lower the mass for a given magnitude. As mentioned by these authors, however, variations of the MF obtained within the afore-mentioned metallicity range remain within the Poisson error bars of the value determined for a solar abundance.

As discussed in detail in BCAH98 (\S 3) and Delfosse et al. (2000), the MMR in infrared bands is very weakly affected by metallicity variations in the afore-mentioned range.
Indeed, metallicity effects for a given mass cancel out with the effect of decreasing effective temperature on the spectral energy distribution in infrared
bands whereas they add up in the V-band.
Therefore, the derivation of the MF from a LF determined {\it directly} in
near-IR bands should be insensitive to biases due to a metallicity spread in the sample. Unfortunately there is presently no direct determination of the LF in such bands. The K-band LF described in \S3 is derived from the V-band LF through various color-color relations and is thus hampered by uncertainties du to these transformations.
The thick-disk population ($\mh \simeq -0.5$), with a scale height $h\approx 0.7$ kpc and a local normalization of $\sim 5\%$ (Haywood, Robin \& Cr\'ez\'e, 1997) should make a negligible contribution to the local sample.

\indent $\bullet$ grain formation : below $\te\approx 2800$ K, grain condensation in the atmosphere of low-mass objects affects drastically their spectral distribution, and thus the mass-magnitude relationship (Chabrier et al., 2000). This corresponds to objects less massive than 0.1 $\msol$ and is not consequential for the present study.

\section{The Galactic disc M-dwarf luminosity-function}

The determination of the M-dwarf LF in the Galactic disk down to the bottom of the MS has
been the source of particular activity within the past recent years.
The determination of the absolute LF requires the determination of the distance of these objects.
The easiest way to determine the distance is by knowing the trigonometric parallax of
the object, which implies a search within near distances from the Sun, typically $d\le 20$ pc for the bright part of the LF ($\mv<9.5$), a few parsecs for the faint end. This
yields the so-called {\it nearby} LF. The main caveat of the nearby LF is that, given the
limited distance, it covers only a limited volume and thus a limited sample of objects. This yields
important statistical indeterminations at large magnitudes ($M_V\simgr 12$).
 On the other hand a fundamental advantage of the nearby LF, besides
the reduced error on the distance, i.e. on the magnitude, is the accurate identification of binary systems.
Other determinations of the disk LF are based on photographic surveys, which extend to
$d\approx 100-200$ pc from the Sun, and thus encompass a significantly larger amount of stars.
However, photometric LFs suffer in general from significant Malmquist bias and, most importantly,
the low spatial resolution of photographic surveys does not allow the resolution of binaries at faint magnitudes.
An extensive analysis of the different nearby and photometric LFs has been conducted by Kroupa
(Kroupa, 1995; see also Reid \& Gizis, 1997, and references therein).
As shown by this author, most of the discrepancy between photometric and nearby LFs
for $M_V>12$ results from Malmquist bias and unresolved binary systems in the low-spatial resolution photographic surveys.

Kroupa (2001) derived a nearby LF, $\Phi_{near}$, by combining Hipparcos parallax data, which 
is essentialy complete for $\mv < 12$ at r=10 pc, and the sample of nearby stars (Dahn et al., 1986) 
with ground-based parallaxes for $\mv>12$ to a completeness distance r=5.2 pc.
Reid and Gizis (1997, RG) extended this determination to a larger volume and determined a nearby LF based on a volume sample within 8 pc. Their local sample is drawn from the most recent version of the Catalogue of Nearby Stars (Gliese \& Jahreiss, 1991). Most of 
the stars in this survey have parallaxes. For late K and M dwarfs, however, trigonometric  
parallaxes are not always available; in such cases, these authors use a spectroscopic TiO-index vs $\mv^{TiO}$ relation 
 to estimate the distance (Reid et al., 1995). This sample was revised recently with Hipparcos measurements 
and new binary detections in the solar neighborhood (Delfosse et al., 1999) and leads to a
revised northern 8-pc catalogue and nearby LF. RG argue that their sample
should be essentially complete for $\mv\le 14$, with all {\it stellar} companions resolved. About 35\% of the systems in the RG sample are multiple and $\sim 45\%$ of all stars have a companion in binary or multiple systems.

The HST LF (Gould, Bahcall \& Flynn, 1997), $\Phi_{HST}$, extends previous photometric surveys to an apparent magnitude $I\simle 24$.
The Malmquist bias is negligible because all stars down to $\sim 0.1\,\msol$ are seen through to the edge
of the thick disk.
A major caveat of any photometric LF, however, is that the determination of the distance relies on a photometric determination from a color-magnitude diagram. In principle this requires the determination of the metallicity of the stars since colors depend on metallicity (see e.g. Figure 12 of Chabrier \& Baraffe, 2000). This is particularly
true for the HST LF since a substantial number of stars in the sample belong to the thick disk and have metal-depleted abundances. Moreover, as mentioned above, the HST small field of view does not allow binary resolution so that
the HST misses essentially all the binaries,
yielding only the determination of the stellar {\it system} LF.
Although, as mentioned above, unresolved binary systems help resolving the differences between the nearby LF and the HST LF, they cannot bring the two LFs into agreement and at least a factor 2 difference remains at the faint end ($\mv > 14$) of the LF and of the MF (see M\'era, Chabrier \& Schaeffer, 1998, their figure 1, with Figures \ref{Fig1_col.ps} and \ref{Fig2_col.ps} below). The source of this disagreement is presently unclear and will be the object of a further study.

The illustration of the disagreement between the three afore-mentioned LFs, $\Phi_{5.2pc}$, $\Phi_{8pc}$ and
$\Phi_{HST}$ in the V-band
at faint magnitude ($\mv\simgr 12$) is shown in Figure 15 of Chabrier \& Baraffe (2000).
Whereas the photometric LF exhibits a pronounced peak around $\mv\approx 12$ ($\mi\approx 10$), with a sharp drop below,
the two nearby LFs, although still peaking at this magnitude, remain rather flattish up to $\mv\sim 15$,
predicting a substantially larger number of stars in this region, a disagreement which cannot be
explained only by unresolved binaries, as mentioned above.
The reasonable agreement between the two nearby LFs, $\Phi_{5.2pc}$ and $\Phi_{8pc}$, down to $\mv\sim 15$, about the
limit of completeness claimed for the 8-pc sample, brings some confidence in these determinations.
The disagreement between these two LFs at fainter magnitude stems very likely from incompleteness of the 8-pc sample.
As shown by Henry et al. (1997; their figure 1) and Delfosse et al. (1999), the M-dwarf sample becomes
substantially incomplete for distances larger than 5 pc.
Second, several M-dwarfs in the RG sample, which represent the faint end population of the LF, lack parallax determination
and have distances estimated from the afore-mentioned spectroscopic relation. This relation can be uncertain by more than 1 magnitude for $\mv \ge 10$ (see Figure 3 of Reid et al., 1995), a major source of Malmquist bias.
On the other hand, the last bins in the $\Phi_{5.2pc}$ LF ($\mv\ge 16$), which include only 5 stars or less each,
might be contaminated by one or several young brown dwarfs, by young pre-MS objects, or by low-metallicity (brighter) very-low-mass stars.
It is unfortunately not possible at the present stage to resolve this issue. We have thus calculated MFs
from both nearby LFs, which
represent the most complete LFs for M-dwarfs presently available.

RG have also derived the LF in the K-band. However, as mentioned above, there is no direct determination of the K-band LF, at least for M-dwarfs, and these authors used $(V-I)$ vs $(V-K$) or $BC_I$ vs $(V-I$) relations, or $M_K$ vs $M_V$ relations for stars with only an absolute magnitude, fitted to the Leggett (1992) data (see below).

\section{The mass function}

Figure \ref{Fig1_col.ps} displays the MF $\xi(\log m)$ calculated with the theoretical MMRs mentioned in \S2 from the V-band LFs $\Phi_{V_{5.2pc}}$ (solid line), $\Phi_{V_{8pc}}$ (dash line) and from the V-band LF converted in $K$ $\Phi_{K_{8pc}}$ (dotted line), in 10$^3$ pc$^3$.
The error bars are the ones quoted by the authors for the RG revised sample, and are
1$\sigma$ Poisson errors for the 5.2 pc sample.
Note that the MF from the RG 8-pc sample has been calculated from a star-by-star analysis, from the sample kindly provided by Neill Reid, not from eqn.(1). We verified, however, that both methods yield similar
results, within the error bars.
The first interesting result is the fact that the MFs derived from the two different V-band LFs agree very well, within 1$\sigma$ in average, down to $\log m=-0.8$ ($m=0.15\,\msol$). The highest mass bin of the 8-pc MF lies several $\sigma$'s above the MF derived from
the Hipparcos sample and must be considered with caution. The 2-$\sigma$ discrepancy between these two MFs below $\log m\sim -0.9$ ($m=0.12$ $\msol$, $\mv\simeq 14.5$ at 5 Gyr) stems very likely from the acknowledged incompleteness of 
$\Phi_{8pc}$ at these magnitudes. On the other hand, the last bin of the 5.2-pc MF may be contaminated by a massive BD or a young pre-MS star, as mentioned previously, and thus should be considered with limited confidence. 
The MF derived from the K-band $\Phi_{8pc}$ LF differs
at the $2\sigma$ level from the two previous ones, and exhibits a more flattish general behaviour.
There is a priori no reason for such a difference since the sample is exactly the same as for the $\Phi_{V_{8pc}}$ LF. There are two possible explanations for such a behaviour:

\indent i) the disagreement might stem from discrepancies in the m-$\mv$ behaviour, as mentioned above. This will be examined in detail in the next subsection.

\indent ii) only 68 of the 106 systems in the 8-pc sample have wide-field K-band imaging.
For objects in the sample lacking IR observations, the $\mk$ absolute magnitude is determined either from a $(V-K)$ versus $(V-I)$ relation or from a $\mk$ versus $\mv$ relations, for those lacking $V,R,I$ photometry, based on the Leggett (1992) red dwarf sample
(see RG for details). Although such relations provide reasonable {\it average} transformations, they do not take into account the color spread in the observational diagram. In particular the $\sim$ 0.2-0.3 mag spread in color in the
observed $(V-K)/(V-I)$ diagram for $2.5\simle V-I \simle 3.5$, i.e. $0.4\simgr m/\msol \simgr 0.2$. This translates into a significant uncertainty in the derived K-band LF. In order to examine this effect, we have recalculated the K-magnitude in the RG sample for objects with $V,R,I$ photometry with a slightly different relation:

\begin{eqnarray}
(V-K)=1.075 + 1.4\times(V-I)
\end{eqnarray}

\noindent which seems to provide a better mean fit of the observed $(V-K)/(V-I)$ relation
than the one used by RG. The MF derived from this new $\Phi_{K_{8pc}}$ is shown in the next subsection.

\subsection{Uncertainty due to the mass-magnitude relation}

As stressed previously, the reliability of the MF depends crucially on the accuracy of the MMR.
As mentioned in \S2 and discussed in BCAH98 and Delfosse et al. (2000), there is a discrepancy between the theoretical mass-$\mv$ relationship derived by BCAH98 and the observationally derived relation below $\sim 0.35\,\msol$, the models being systematically brighter than the observations, with a maximum $\sim 0.5$ magnitude discrepancy around $m=0.2\,\msol$. 
This was anticipated from an identified shortcoming of the present theory in the V-band (see BCAH98), mainly due to still incomplete molecular linelists and thus opacity coefficients in the optical. Therefore, the models underestimate the mass for a given magnitude below $\sim 0.35\,\msol$. The effect, however, remains modest: for example a magnitude $\mv=12$ corresponds to an observed mass $m\simeq 0.28\,\msol$, whereas the theory predicts $m\simeq 0.25\,\msol$; the maximum disagreement around $\mv=13$ corresponds to an observed mass $m\simeq 0.2\,\msol$, whereas the theory predicts $m\simeq 0.17\,\msol$. In order to quantify the effect on the MF, we have recalculated it with the parametrized mass-$\mv$ relation of Delfosse et al. (2000) which provides a good mean fit of their data. This polynomial fit reads (see Delfosse et al., 2000):

\begin{eqnarray}
\log (m/\msol)=10^{-3}\times [0.3+1.87\times\mv + 7.614\times \mv^2
-1.698\times \mv^3 + 0.060958\times \mv^4]\\
\nonumber
\,\,\,{\rm for}\,\,\mv \in [9,17]
\end{eqnarray}

Figure \ref{Fig2_col.ps} displays the MF derived from $\Phi_{V_{5.2pc}}$ and $\Phi_{V_{8pc}}$ when using this relation, to be compared with Figure \ref{Fig1_col.ps}.
As anticipated from the discussion above, the main effect of the Delfosse et al. (2000) $m-\mv$ relation is to increase the number of stars in the mass range $m\sim 0.2-0.4\,\msol$ by slightly depressing both the higher mass tail and the very-low-mass tail of the distribution, yielding a slightly more flattish distribution below $\log m\approx -0.5$ ($m\approx 0.3\,\msol$).
Figure \ref{Fig2_col.ps} displays also the MF derived from the revised $\Phi_{K_{8pc}}$ LF, obtained with eqn.(4), as described above. Interestingly, the three MFs seem now to agree reasonably well. The 1 to 2 $\sigma$ difference between the MF derived from the 5.2-pc sample and the ones derived from the 8-pc sample for $\log m \simle -0.8$ stems very likely from (i) incompleteness of this latter at faint magnitudes, (ii) remaining uncertainties in the $\mv$ into $\mk$ transformation for $\Phi_{K_{8pc}}$, and (iii) from the presence of BD or PMS objects in the last bin of $\Phi_{V_{5.2pc}}$. Because of point (ii), and until
the M-dwarf LF is {\it observationally}-determined in infrared filters, we will focus in the following on the MF derived from the two observed V-band LFs.

\subsubsection{Parametrization of the mass fonction}

Figure \ref{Fig3_color.ps} displays the two MFs derived from the V-band LFs
with three different functional forms superposed, namely:

\indent- a power-law form (solid line; IMF1):

\begin{eqnarray}
\xi(m)={dn\over dm}=A\,m^{-\alpha}
\end{eqnarray}

\indent- a log-normal form (dot-dash line; IMF2):

\begin{eqnarray}
\xi(\log \,m)={dn\over d\log \, m}=A\,exp\{-{(\log \, m\,\,-\,\,\log \, m_0)^2\over 2\,\sigma^2}\}
\end{eqnarray}

\indent- an exponential form (dash line; IMF3):

\begin{eqnarray}
\xi(m)={dn\over dm}=
A\,m^{-\alpha}\, exp\{-({ m_0\over m})^{\beta}\}
\end{eqnarray}

This latter form resembles a log-normal form in the low-mass range and recovers asymptotically to
a power-law at large $m$, $\xi(m)_{m>>m_0}\rightarrow m^{-\alpha}$.
The peak value corresponds to $m_p=({\beta \over \alpha})^{(1/\beta)}m_0$.
Such a functional form was proposed originally by Larson (1986) in the case of a bimodal star formation process in galactic disks and a remnant dominated dark matter component in the solar neighborhood.
It has been recently advocated as the IMF required for a dark halo containing a substantial population of
stellar remnants, accounting for the microlensing observations towards the LMC (Chabrier et al., 1996; Chabrier 1999).

All three functional forms are normalized at 0.8 $\msol$ on the Hipparcos value $({dn\over dm})_{0.8\,\msol}=
2.8\pm 0.2\,\times 10^{-2}\mvol$.
Stars below $m\simeq 0.8\,\msol$ did not have time to evolve off the MS within $\sim 10$ Gyr, about the age of the Galactic disk, so the present-day MF for stars below 0.8 $\msol$ reflects the IMF.
One of the constraints to be fulfilled by the IMF is the so-called {\it continuity constraint}, which requires
the IMF to be continuous at the mass for which the main sequence lifetime $t_{MS}$ corresponds to the age of the disk
$t_{MS}\approx t_{D}$ (Miller \& Scalo, 1979). This mass is about $\sim 0.8$-$0.9$ $\msol$. For larger masses, the
IMF seems to be well described by a power-law with an exponent $\alpha\simeq 2.7$ (Scalo, 1986), although substantial uncertainties remain in the exact determination on the exponent. Therefore, the forms (6)-(8) must be close to such a mass distribution for masses above 1 $\msol$ to
fulfill the continuity constraint. A condition reasonably well satisfied in all three cases, as shown in Figure \ref{Fig4_color.ps}.  
Table 1 gives the values of the different parameters entering equations (6)-(8)
as well as
the corresponding average mass of star formation:

\begin{eqnarray}
\langle m\rangle={ \int_{m_{\star_{inf}}}^{m_{\star_{sup}}} \xi(m)\,m\,dm \over \int_{m_{\star_{inf}}}^{m_{\star_{sup}}} \xi(m)\,dm}
\end{eqnarray}

\noindent where $m_{\star_{inf}}$ is the stellar minimum mass, by definition the hydrogen-burning limit, $\sim 0.07\, \msol$ for solar-metallicity stars (Chabrier \& Baraffe, 1997), and $m_{\star_{sup}}$ is the maximum mass for star formation, $\sim 100\,\msol$.

As seen in Figure \ref{Fig3_color.ps}, the three analytical forms (6)-(8) provide very reasonable fits
- within less than 2$\sigma$ - of the MF, if the last bin of the MF derived from $\Phi_{5.2pc}$ is omitted\footnote{Taking this last bin into account yields a significantly steeper MF, with $\alpha\simeq 1.8$, as derived in the previous analysis of M\'era, Chabrier \& Baraffe (1996).}. The dotted line displays the
4-segment power-law IMF derived by Kroupa (2001).

Figure \ref{Fig4_color.ps} displays the three MFs over the entire stellar mass-range from 60 $\msol$ down to the
H-burning limit, and the extension in the substellar domain. As seen in the figure, the
exponential IMF is an excellent description of the stellar distribution over the entire characteristic mass
range, recovering a power-law tail at large masses and the log-normal form below $\sim 1\,\msol$. Such a functional form, proposed for the dark halo IMF, as mentioned above, is thus not an ad-hoc prescription but
represents quite well the Galactic disk stellar IMF. Width and normalization are of course different for the
disk and the primordial halo (see Chabrier, 1999), as anticipated from the very different characteristic
conditions (virial temperature, metallicity, etc...).
As seen in the figure, the power-law IMF differs significantly from the two other forms in the substellar regime.
Brown dwarf counts should then enable us to distinguish between these forms. Substellar objects, however,
never reach thermal equilibrium, which defines the MS, and the age distribution must be taken into account when determining
the IMF from the present-day observed distribution. This issue will be addressed in a forthcoming paper.
The log-normal and the exponential forms are very similar: at $\sim 0.01\,\msol$ ($\,\simeq 10\,$ Jupiter mass), the predicted number of objects differs only by about 60\%.
Such a difference
is unlikely to be detectable for objects in the field, with resolved companions.
The IMF derived by Kroupa (2001) is shown for comparison by the dotted line.

Figure \ref{Fig5_color.ps} displays the same MFs below 0.6 $\msol$ in a linear scale. A Salpeter IMF with the same normalization is shown for comparison (dotted line). Such an IMF all the way down to the bottom of the MS would overestimate the stellar
 density at
0.1 $\msol$ by about a factor of 6.

\section{The Galactic disk stellar mass budget}

Integration of the MFs derived in \S4 from 1.0 $\msol$ down to the H-burning minimum mass 0.07 $\msol$ yields the LMS (generically designed here as stars below a solar mass, $m\le 1.0\,\msol$) number-density
$n_{LMS}$ and mass-density $\rho_{LMS}$:

\begin{eqnarray}
\nonumber n_{LMS}=\int_{0.07}^{1.0} \xi(m)\,dm=\int_{\log(0.07)}^{\log(1.0)} \xi(\log m)\,d\log m\\
\rho_{LMS}=\int_{0.07}^{1.0} \xi(m)\,m\,dm=\int_{\log(0.07)}^{\log(1.0)} \xi(\log m)\,m\,d\log m
\end{eqnarray}

Table 2 gives $n_{LMS}$ and $\rho_{LMS}$ for the
three analytical MFs given by eqns. (6)-(8).
Adding the more massive stars with the scale height dependent present-day MF (PDMF) of Scalo (1986, Table 4) yields
the present-day disk main sequence stellar density. Adding up the {\it single} white dwarf density $n_{WD}\simeq 5.5\pm 0.8 \,\times 10^{-3}$ pc$^{-3}$ (Holberg, Oswalt \& Sion, 2001)
with an average mass $\langle m_{WD}\rangle =0.67\,\msol$, i.e.
a white dwarf mass density $\rho_{WD}\simeq 3.7\pm 0.5 \,\times 10^{-3}\,\mvol$,
a neutron star density $n_{NS}\simeq 10^{-3}$ pc$^{-3}$ (Popov et al., 2000) with a mass $\langle m_{NS}\rangle=1.4\,\msol$ and a red giant contribution $n_{RG}\simeq 0.3\times 10^{-3}$ pc$^{-3}$, $\rho_{RG}\simeq 0.6\times 10^{-3}\,\mvol$ (Haywood et al., 1997) yields the present-day
Galactic thin-disk total stellar number- and mass-densities $n_\star\simeq 0.13\times 10^{-2}$ pc$^{-3}$, $\rho_\star\simeq 4.30\pm 0.3 \,\times 10^{-2}\,\mvol$.
Assuming a standard double-exponential thin-disk+thick-disk with a common scale length $l=2.5$ kpc and
scale heights $h=250$ pc and $h\simeq 760$-1000 pc, respectively, for stars with $m\le 1.0 \,\msol$, and a Scalo (1986) mass-dependent scale height for more massive stars in the thin-disk, with a thick-disk local normalization $\sim 5\%$ (Haywood et al., 1997), we obtain the stellar contribution to the disk
local surface density $\Sigma_{\odot_\star}\simeq 24.0\pm 2$ $\msol$ pc$^{-2}$.
The $\sim 10\%$ error bars on $\rho_{LMS}$ correspond to the 1$\sigma$ Poisson uncertainty on the MF at $\log m=-0.9$. These numbers are still affected by the uncertainty at the very-low-mass end of the stellar MF, due to small statistics and incompleteness, and by the uncertainty in the exact scale height and normalization of the thick disk. These uncertainties, however, should not affect the present numbers by more than $\sim 10\%$.
An accurate determination of the MF below $m\sim 0.15\,\msol$, i.e of the LF at $\mv \simgr 14$, $\mk \simgr 9$, is strongly needed to reduce this uncertainty
and to nail down the normalization of the stellar MF at the H-burning limit.

\section{Conclusion}
 
We have derived the low-mass star mass-function characteristic of the disk
stellar population from parallax-determined LFs down to the vicinity of the hydrogen-burning limit.
This stellar MF is derived from a
consistent stellar evolution theory which accurately describes the very mechanical and thermal properties of these stars, and yields reliable mass-magnitude relationships.
The consequences of the remaining uncertainties in the theoretical mass-$\mv$ relationship on the determination of the MF have been examined carefully. The MFs derived from both the 5.2-pc and 8-pc LFs in the V-band agree fairly well ($<1\,\sigma$) down to $\sim 0.15\,\msol$. The MF derived from the K-band LF is in reasonable agreement with the previous ones, but its accuracy is hampered by  photometric or magnitude transformations from $\mk$ to $\mv$ based on simple mean fits. The direct determination of the LF in the K-band would be of tremendous interest, since 
the dispersion in the mass-magnitude relation due to the metallicity spread is negligible in this band. Incompletude is likely to affect the 8-pc sample at magnitudes corresponding to the above-mentioned mass.
Age effects are consequential only for objects below $\sim 0.1\,\msol$ and
younger than $\sim 0.5$ Gyr and thus do not affect significantly the main sequence stellar mass function above this limit.
This can affect, however, the very faint end of the nearby 5.2-pc LF, leading
to the misidentification of a statistically significant number of young brown dwarfs or pre-MS very-low-mass stars as main sequence low-mass stars.
Nevertheless, agreement between the various nearby
surveys down to $\mv\sim 15$ brings confidence in the present determination of the stellar MF down to about 0.15 $\msol$. 

The MF is found to flatten significantly below about $\sim 1.0\,\msol$, compared with a Salpeter MF, and again below $\sim 0.5 \,\msol$, as noted originally by Miller \& Scalo (1979) and Kroupa et al. (1993), but keeps rising slowly down to the bottom of the main sequence.
Incompleteness and limited statistics prevent the unambiguous determination of the value of the mass function at the very hydrogen-burning limit.
Larger statistics is needed at faint magnitudes, a difficult task since parallax-based surveys are likely
to be affected by incompleteness beyond $\sim 5$ pc and photometric surveys lack binary resolution, not mentioning possible erros due to Malmquiest bias or color-magnitude transformations.

The IMF is well
described over the entire stellar mass range, i.e. four orders of magnitude in mass,
by three different functional forms, namely a two-segment power-law, a log-normal and an exponential.
Interestingly, this latter form, which has been advocated recently to suggest a substantial population of old stellar remnants in the dark
halo, seems to offer the best compromise between the high-mass tail and the low-mass end of the stellar distribution.
Only in the substellar regime does the power-law IMF differ significantly from the two other forms, predicting
a substantially larger number of field brown dwarfs. This issue will be addressed in a forthcoming paper.
The previously suggested bimodal behaviour for the IMF, with a maximum around
$\sim 0.3 M_\odot$ and a drop below (Scalo, 1986), seems to be excluded. It is very likely
an artifact due to (i) unresolved binaries
in the photometric luminosity functions (see Kroupa et al., 1993), (ii) small statistics in the nearby LF, (iii) the use of inaccurate mass-magnitude relationships
in the M-dwarf regime.

Integrations of this IMF below 1 $\msol$ and of the present-day MF for main sequence stars above 1 $\msol$, plus the contribution of stellar remnants and red giants, yield the
presently most reliable determination of the disk stellar density and its contribution to the Galactic disk mass budget.

\bigskip

\begin{acknowledgements} The author is indebted to Pavel Kroupa and Neill Reid for providing their luminosity functions and for various discussions.
The author is most grateful to the Astronomy Department and to the Miller Institute of the University of Berkeley, where most of this work was conducted, for a visiting professor position and for their warm hospitality.
\end{acknowledgements}

\clearpage\eject

\bigskip
\begin{table}
\caption[]{Parameters of the mass functions}
\bigskip
\begin{tabular}{lcccc}
\tableline
& $\xi(m)$ (eqn.(6),IMF1) & $\xi(\log \,m)$ (eqn.(7),IMF2) & $\xi(m)$ (eqn.(8),IMF3)  \\
\hline \\
A       & 0.019 M$_\odot^{-1}$ pc$^{-3}$ & 0.141 pc$^{-3}$ & 3.0 M$_\odot^{-1}$ pc$^{-3}$  \\
$\alpha$  &                             &                              & 3.3  \\
\mbox{}\hspace{0.3cm}  $m\le 1.0\,\msol$ & 1.55  &                     &      \\
\mbox{}\hspace{0.3cm}  $m> 1.0\,\msol$ & 2.70  &                     &      \\
$m_0$ ($\,\msol$)  &                     & 0.1                         & 716.4\\
$\beta $  &                              &                             & 0.25 \\
$\sigma$  &                              & 0.627                       &      \\
\hline \\
$\langle m \rangle$  ($\,\msol$) & 0.44   & 0.45                       & 0.55 \\
\tableline
\end{tabular}
\end{table}

\newpage

\vfill\eject

\bigskip

\begin{table}
\caption[]{Disk present-day stellar density.\label{table2}}
\bigskip
\begin{tabular}{lcccc}
\tableline
& scale height & $n_\star$ & $\rho_\star$ & $\Sigma_\odot$   \\
& (pc) & (pc$^{-3}$) & (M$_\odot$ pc$^{-3}$) & (M$_\odot$ pc$^{-2}$) \\
\hline \\
{\bf Thin disk}:  &  &  &  &   \\
\mbox{}\hspace{0.5cm} LMS ($\le 1.0\,\msol)$      & 250 & 0.12$\pm$ 0.02 & $3.10\pm 0.3 \,\times 10^{-2}$  & $15.5\pm 2$ \\
\mbox{}\hspace{0.5cm}  MS stars $> 1.0\,\msol$     & Scalo'86 &$0.43\times 10^{-2}$ &$0.6\times 10^{-2}$ & 2.3 \\
\mbox{}\hspace{0.5cm}  WD + NS + RG & 250 & $0.7\times 10^{-2}$ &$0.6\times 10^{-2}$ & 2.8 \\
\mbox{}\hspace{0.5cm}  all stars    &  & $0.13\pm 0.02$ & $4.30\pm 0.3 \,\times 10^{-2}$ & $20.6\pm 2$ \\
{\bf Thick disk}:   & 760-1000 &  & &  \\
\mbox{}\hspace{0.5cm}  all stars    & & & $\approx 0.22\pm 0.02 \times 10^{-2}$ & $\approx 3.2-4.3$ \\
\hline \\
{\bf Total}  &  & $0.13\pm 0.02$ & $4.50\pm 0.3 \,\times 10^{-2}$ & $24.4 \pm 2.5$\\
\tableline
\end{tabular}
\end{table}

\newpage

\vfill

\clearpage\eject

\begin{figure}

\centerline {\bf FIGURE CAPTIONS}
\vskip1cm

\caption[]{Mass functions $\log [\xi (\log m)]$ in 10$^3$ pc$^3$ derived from the LFs
$\Phi_{V_{5.2pc}}$ (solid line), $\Phi_{V_{8pc}}$ (dashed line) and
 $\Phi_{K_{8pc}}$ (dotted line), with the BCAH (1998) m-$\mv$ and m-$\mk$ relationships.}
\label{Fig1_col.ps}
\end{figure}

\begin{figure}
\caption[]{Same as Fig. \ref{Fig1_col.ps} but calculated with the Delfosse et al. (2000)
 m-$\mv$ relationship for $m\le 0.6\,\msol$ and from the corrected $\Phi_{K_{8pc}}$ (see text).}
\label{Fig2_col.ps}
\end{figure}

\begin{figure}
\caption[]{Same as Fig. \ref{Fig2_col.ps} for the MFs derived from $\Phi_{V_{5.2pc}}$ (solid line) and $\Phi_{V_{8pc}}$ (dashed line). The curves display the 3 different functional forms: power-law (IMF1; solid), log-normal (IMF2; dash-dot) and exponential (IMF3; dash) (see text and Table 1), all normalized at 0.8 $\msol$. The Kroupa (2001) IMF is shown by the dotted line, as a series of power-laws.}
\label{Fig3_color.ps}
\end{figure}

\begin{figure}
\caption[]{Same as Fig. \ref{Fig3_color.ps} over the mass-range $0.01\,\msol \le m \le 60\,\msol$.}
\label{Fig4_color.ps}
\end{figure}

\begin{figure}
\caption[]{Same as Fig. \ref{Fig3_color.ps} in a linear scale $\xi(m)$ over the mass-range $0.1\,\msol \le m \le 0.6\,\msol$. The dotted line corresponds to a Salpeter IMF with the same normalization.}
\label{Fig5_color.ps}
\end{figure}

\vfill\eject

\vfill\eject
\begin{figure}
\begin{center}
\epsfxsize=190mm
\epsfysize=200mm
\epsfbox{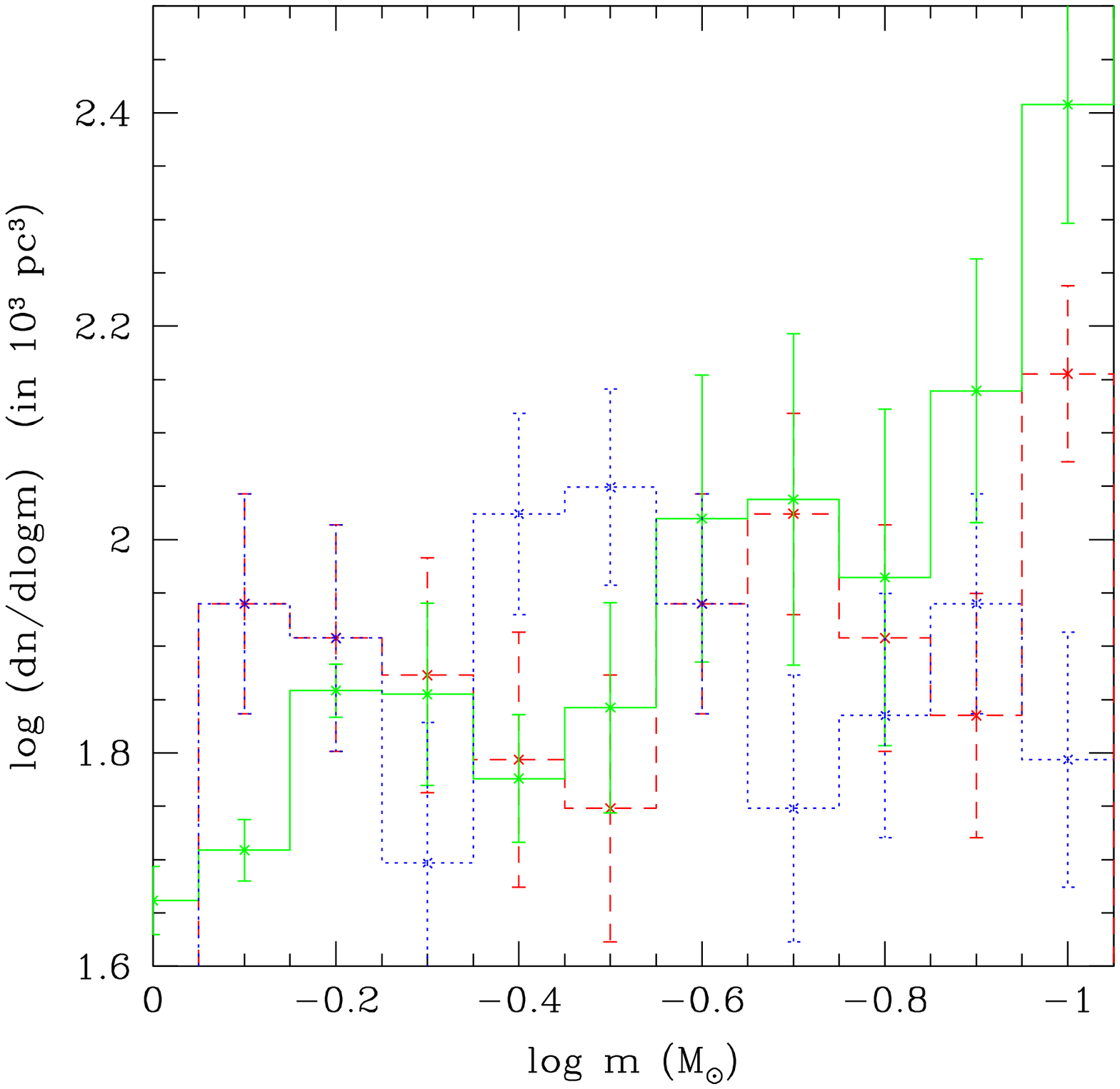}
\end{center}
\end{figure}

\vfill\eject

\vfill\eject
\begin{figure}
\begin{center}
\epsfxsize=190mm
\epsfysize=200mm
\epsfbox{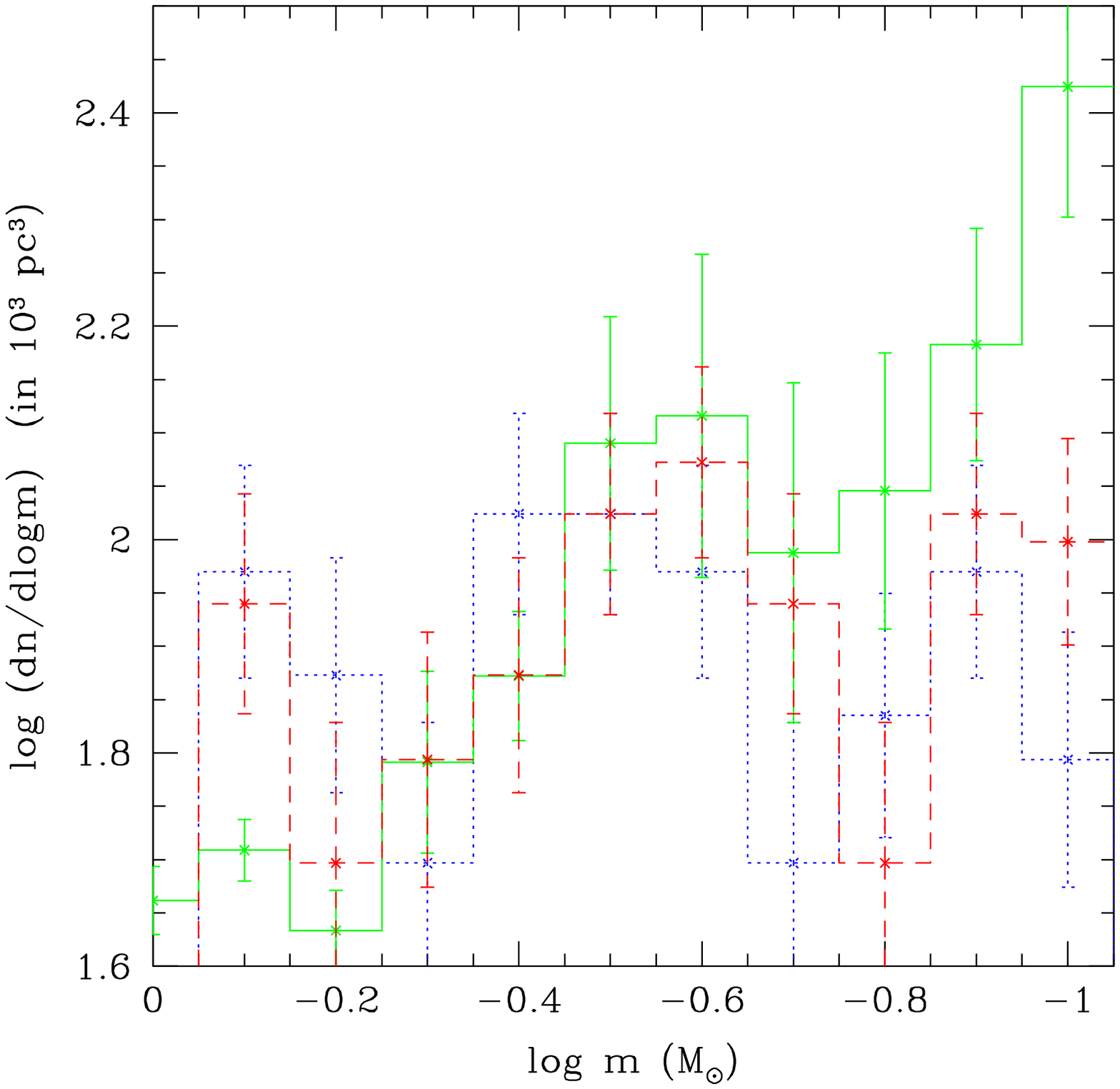}
\end{center}
\end{figure}

\vfill\eject

\vfill\eject
\begin{figure}
\begin{center}
\epsfxsize=190mm
\epsfysize=200mm
\epsfbox{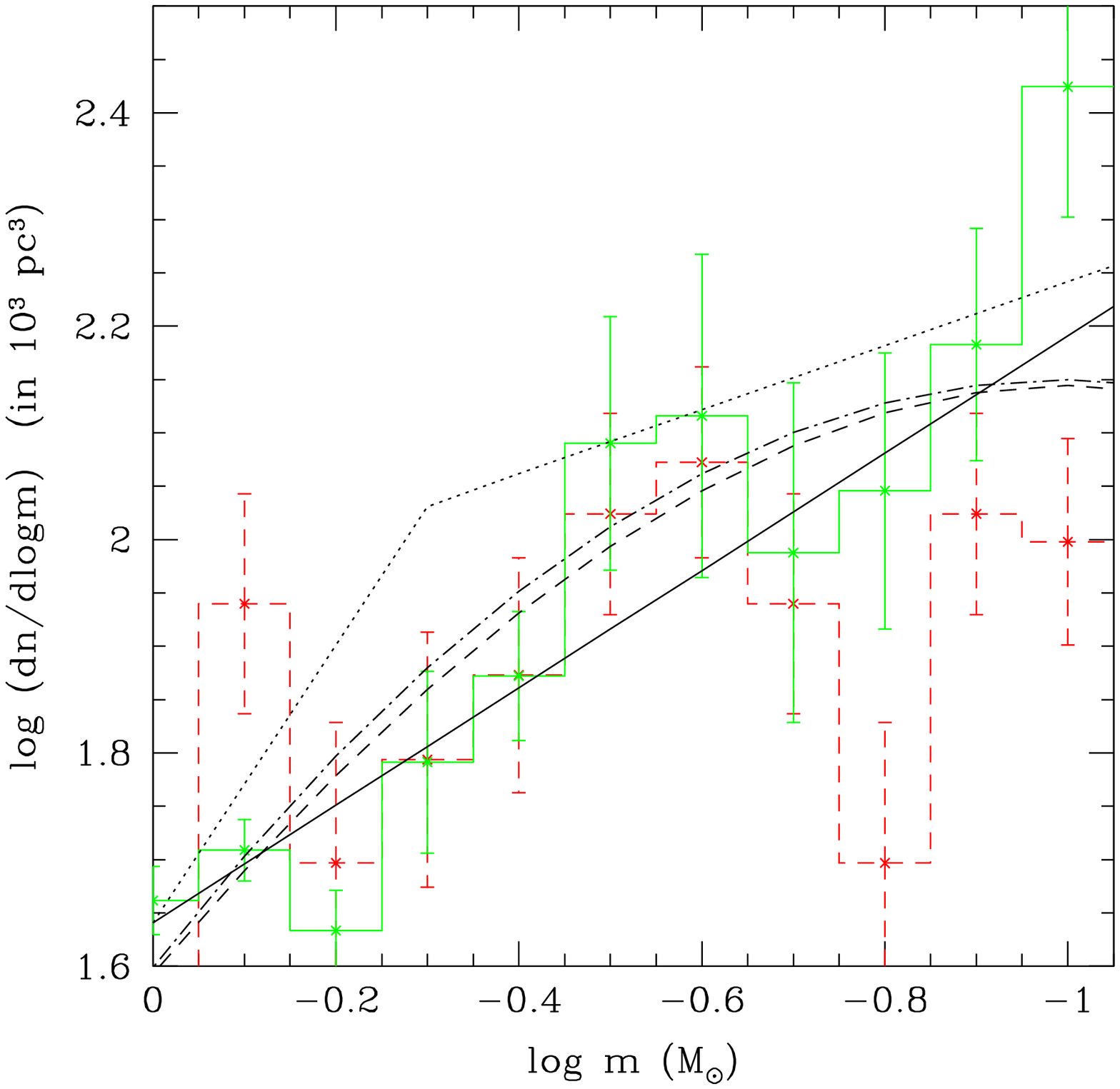}
\end{center}
\end{figure}

\vfill\eject

\vfill\eject
\begin{figure}
\begin{center}
\epsfxsize=190mm
\epsfysize=200mm
\epsfbox{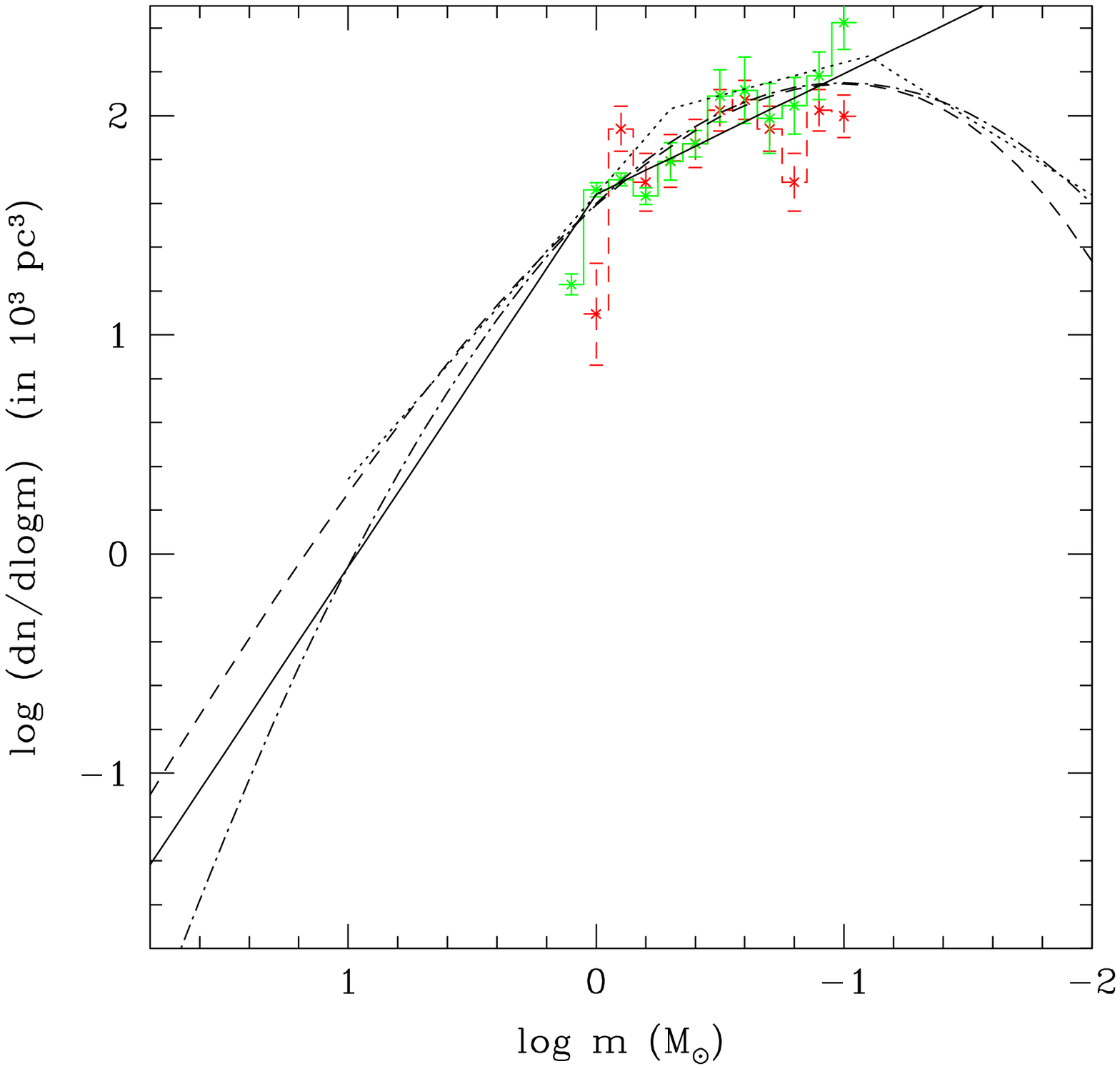}
\end{center}
\end{figure}

\vfill\eject

\vfill\eject
\begin{figure}
\begin{center}
\epsfxsize=190mm
\epsfysize=200mm
\epsfbox{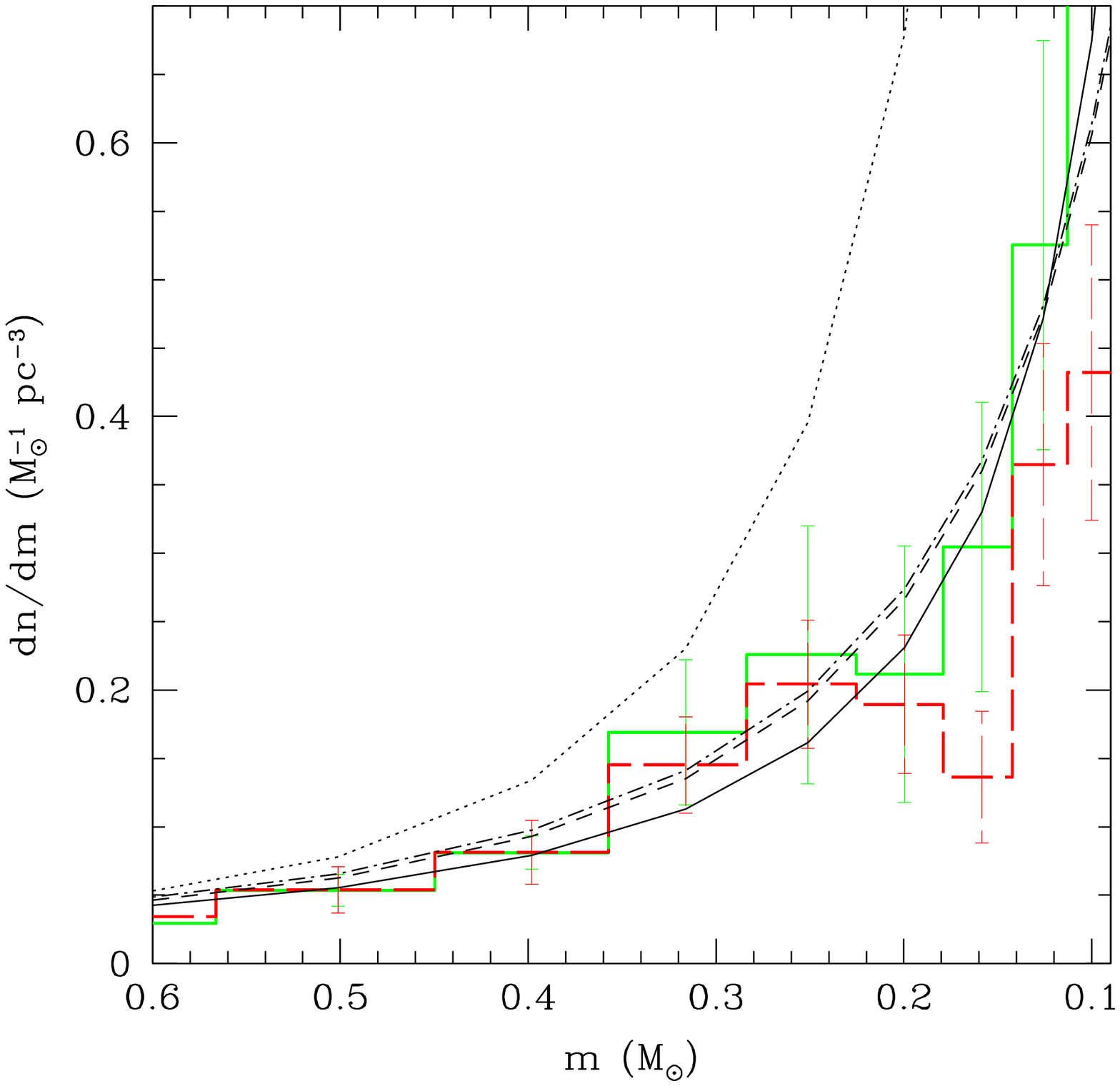}
\end{center}
\end{figure}

\vfill\eject

\end{document}